\documentclass[conference, 10pt]{IEEEtran}

\usepackage{amsfonts}
\usepackage{amsthm}
\usepackage{xcolor,soul,framed} 
\usepackage[bookmarks=false]{hyperref}
\usepackage[left=1.62cm,right=1.62cm,top=1.9cm, bottom=1in]{geometry}

\colorlet{shadecolor}{yellow}
\usepackage[pdftex]{graphicx}
\graphicspath{{../pdf/}{../jpeg/}}
\DeclareGraphicsExtensions{.pdf,.jpeg,.png}

\usepackage[cmex10]{amsmath}

\usepackage{array}
\usepackage{subfig} 
\usepackage{mdwmath}
\usepackage{amsmath}
\usepackage{amssymb}
\usepackage{mathtools}
\usepackage{mathrsfs}
\usepackage{mdwtab}
\usepackage{eqparbox}
\usepackage{url}
\usepackage{blkarray}
\usepackage{newtxtext,newtxmath}

\usepackage{atbegshi} 
\newlength{\origcolumnsep}

\usepackage[binary-units,per-mode=symbol]{siunitx}
\usepackage[linesnumbered,ruled,vlined,noend]{algorithm2e}

\usepackage{xcolor}

\SetCommentSty{mycommfont}
\SetKwInput{KwInput}{Input}
\SetKwInput{KwOutput}{Output}
\SetKwInput{KwRequire}{Require}
\SetKwInput{KwEnsure}{Ensure}
\SetKw{KwRet}{return}

\usepackage{booktabs}

\usepackage{algpseudocode} 
\usepackage{cuted}
\usepackage{balance} 
\usepackage{etoolbox}

\begin{document}
 \bstctlcite{IEEEexample:BSTcontrol}
     \title{Quota-Based Aging for Prediction-Guided Scheduling in LLM Serving}


\newcommand{\timothy}[1]{\textcolor{cyan}{[#1]}}
\newcommand{\qiyue}[1]{\textcolor{magenta}{[#1]}}


\author{
    \IEEEauthorblockN{Hanzhi Zhang, Anxuan Che, Hang Qi, Qiyue Li, Jingjin Wu}
    \IEEEauthorblockA{Guangdong Prov./Zhuhai Key Lab of IRADS, Beijing Normal-Hong Kong Baptist University,
    Guangdong, P. R. China\\
    Email: t330005063@mail.bnbu.edu.cn; t330018003@mail.bnbu.edu.cn; t330202706@mail.bnbu.edu.cn;\\
    t330005026@mail.bnbu.edu.cn; jj.wu@ieee.org\\
    }
}
\maketitle

\begin{abstract}
Large language model (LLM) serving systems use token-level
continuous batching to improve resource utilization, but heterogeneous
response lengths complicate request scheduling. Prediction-guided
short-first scheduling reduces average latency but can severely delay
long requests, while unrestricted aging may over-promote delayed long
requests under congestion. We propose Quota-Based Aging (QBA), a
lightweight policy that combines predicted short-first admission with
quota-bounded, debt-enforced promotion of aged predicted-long requests.
We also use a reduced M/G/$n$ queueing surrogate, calibrated from
simulator traces, to explain first-order mean-latency trends. QBA is
evaluated using a serving-aware event-driven simulator that models
decode-priority batching, chunked prefill, token-budget constraints,
and KV-cache-safe admission. We compare QBA with FCFS, predicted
short-first, and aggressive aging under asymmetric prediction errors
on synthetic and production-derived workloads. The evaluation considers
queueing delay, throughput, utilization, and class-conditional tail
performance. Results show that QBA provides a controllable middle
ground between aggressive short-first prioritization and unrestricted
aging, substantially reducing long-request tail latency while retaining
much of the mean-latency benefit of prediction-guided scheduling.
\end{abstract}

\begin{IEEEkeywords}
large language model serving, request scheduling, Quality of Service, aging mechanism, queueing theory
\end{IEEEkeywords}

%
\IEEEpeerreviewmaketitle


\section{Introduction}
Large Language Models (LLMs) support interactive services such as dialogue, coding, and content generation~\cite{brown2020language,openai2023gpt4}. Once deployed as online services, they must process dynamically arriving requests under stringent latency requirements. Modern serving engines improve resource utilization through token-level continuous batching~\cite{zhang_edge_llm_batching_2024,kwon_pagedattention_2023}. However, this execution model couples scheduling to heterogeneous prompt and response lengths, competing prefill and decode stages, a finite token budget, and KV-cache capacity. Scheduling therefore becomes a core performance-control mechanism: it affects not only average responsiveness, but also whether particular request classes experience extreme delay. Accordingly, we use Quality of Service (QoS) here to refer to the joint control of mean latency and class-conditional tail latency. 


Length prediction provides a natural signal for exploiting request heterogeneity. Prioritizing requests predicted to be short can reduce mean delay, consistent with classical short-job-oriented scheduling~\cite{schrage_srpt_1966,harchol_balter_2013}, while recent LLM serving work has explored latency- and efficiency-oriented execution mechanisms~\cite{wu2023llm,sheng2023flexgen}. However, predicted short-first (PSF) admission can repeatedly defer predicted-long requests and inflate their tail latency, especially under congestion. Generic aging can mitigate starvation~\cite{silberschatz2018operating}, but unrestricted promotion may place too many delayed long requests ahead of short ones and erase the mean-delay benefit that motivated prediction-guided scheduling. Accordingly, we ask the following research question: \textbf{How can an LLM serving scheduler use coarse binary response-length predictions to retain the mean-delay benefit of short-first admission while controlling long-request tail inflation, without aging-based promotions overwhelming the short-first path under token-level serving constraints?}


To answer this question, we propose \emph{Quota-Based Aging (QBA)}, a QoS-aware bounded-aging scheduler for prediction-guided LLM serving. QBA prioritizes predicted-short requests while allowing delayed predicted-long requests to become eligible for promotion through a quantile-based waiting threshold. Rather than promoting every aged predicted-long request, QBA earns one promotion opportunity after every $k$ successful short admissions. This opportunity is recorded as promotion debt and is cleared only when an aged predicted-long request is actually admitted. By separating aging eligibility from promotion frequency, QBA provides bounded but enforceable protection for delayed long requests, with $k$ serving as an interpretable control over the mean–tail latency trade-off.

To characterize this trade-off, we further pair QBA with a reduced M/G/$n$ queueing abstraction and a realistic event-driven simulator. The analytical model intentionally collapses implementation details into effective service and parallelism parameters, making the main delay trends interpretable. The simulator then restores practical serving mechanisms, including chunk continuation, KV-cache-safe admission, and token-load-dependent batch service, so that the analytical observations can be checked under realistic dynamics. We additionally test the same scheduling mechanisms, without re-tuning, on production LLM inference traces to examine whether the observed trade-off transfers beyond the synthetic setting.

The main contributions are:
\begin{itemize}
\item We propose QBA, a practical QoS-aware bounded-aging policy designed for prediction-guided LLM serving. To the best of our knowledge, it is a novel policy targeting the coupled serving features above. QBA separates aging eligibility from actual admission: delayed long requests become eligible through a quantile-based waiting threshold, while promotion is bounded by a quota cycle and enforced through promotion debt.
\item We develop a tractable M/G/$n$ abstraction that explains how load, service variability, and bounded promotion affect mean delay under token-level continuous batching. By collapsing token-level execution into effective service moments and effective parallelism, the model gives a compact view of why short-first scheduling lowers mean delay, why aggressive aging can over-correct, and why QBA provides a more controlled operating point.
\item We validate the analytical trends using realistic event-driven simulation, showing consistent favorable mean-delay/tail-latency trade-off for QBA. The simulator restores serving details absent from the reduced model, including decode-priority batching, continuation handling, KV-cache-safe admission, token-budget constraints, and token-load-dependent batch service.
\item We further validate QBA on 100 non-overlapping windows of a
production Microsoft Azure LLM inference trace ~\cite{patel_splitwise_2024}.  Despite bursty
non-Poisson arrivals and a prompt-response composition substantially
different from the synthetic workload, the results confirm
that QBA remains a
stable intermediate operating point without any workload-specific
re-tuning.
\end{itemize}
The implementation and frozen experimental artifacts are publicly
available at~\cite{qba2026code}.

\begin{figure}[htbp]
    \centering
    \includegraphics[width=\columnwidth]{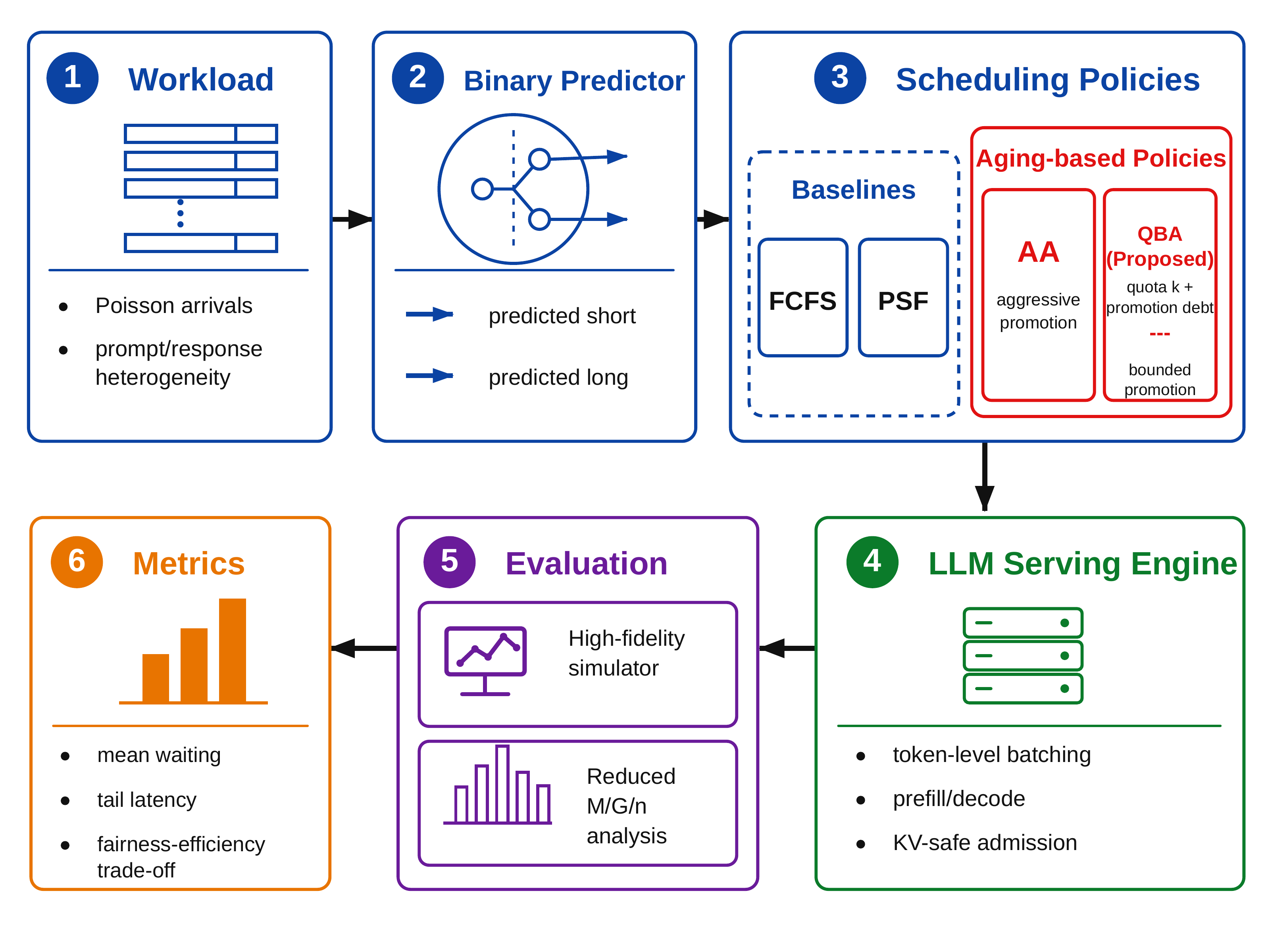}
    \caption{Workflow of prediction-guided LLM request scheduling, including
workload generation, binary response-length prediction, scheduling,
token-level continuous batching, evaluation, and latency metrics.
The synthetic workload follows Poisson arrivals, while the
production trace (Section~IV-E) retains its original bursty arrival
structure.}
    \label{fig:llm_scheduling_workflow}
\end{figure}

\section{System Model}

\begin{table}[t]
\centering
\caption{Key notation.}
\label{tab:notation}
\scriptsize
\renewcommand{\arraystretch}{1.0}
\begin{tabular}{@{}l@{\ \ }l@{\quad}l@{\ \ }l@{}}
\toprule
Sym & Meaning & Sym & Meaning \\
\midrule
$\lambda,\rho$  & arrival rate / load         & $n$             & effective parallelism \\
$P,D,S$         & prompt / decode / eff. svc. & $\tau_{0.85}$        & short/long cutoff \\
$Y_i,\hat Y_i$& true/predicted class        & $\pi_{S/L}$     & class priors \\
$\epsilon_{SL/LS}$ & misclassif. rates       & $\gamma$        & aging quantile level \\
$k$             & QBA quota                   & $\phi_X(\rho)$  & promoted long frac. \\
\bottomrule
\end{tabular}
\end{table}

As summarized in Fig.~\ref{fig:llm_scheduling_workflow}, this section defines the LLM serving system, the reduced queueing abstraction, and the prediction labels used by the scheduler. 

\subsection{LLM Serving Model}

We consider a single-engine LLM serving system with token-level continuous batching~\cite{zhang_edge_llm_batching_2024,kwon_pagedattention_2023}.
For analytical tractability, we first consider that requests arrive according to a Poisson process with rate $\lambda$, a standard open-system abstraction in queueing models~\cite{harchol_balter_2013}.
Request $i$ is represented by $\mathcal R_i=(P_i,D_i,\hat Y_i)$, where $P_i$ is the prompt length, $D_i$ is the response length, and $\hat Y_i\in\{S,L\}$ is a predicted short/long label.

\subsubsection{Two-stage Execution}
Each request evolves through a \emph{prefill} stage followed by a \emph{decode} stage.
Prefill processes prompt tokens and builds the KV cache, while decode generates response tokens iteratively until completion~\cite{patel_splitwise_2024}.
Let $P_i^{\rm rem}(t)$ and $D_i^{\rm rem}(t)$ denote the remaining prefill and decode workloads.
A request departs when $P_i^{\rm rem}(t)=0$ and $D_i^{\rm rem}(t)=0$.

\subsubsection{Decode-Priority Token Batching}
At batch boundary $\ell$, the scheduler forms a batch under token budget $B$, following the batching abstraction used in modern inference-serving systems~\cite{zhang_edge_llm_batching_2024,liu_edge_inference_batching_2023}.
Let $\mathcal D_\ell^\pi$ and $\mathcal P_\ell^\pi$ denote the decode and prefill requests selected by policy $\pi$.
The batch must satisfy
\[
\sum_{i\in\mathcal D_{\ell}^\pi} q_i + \sum_{j\in\mathcal P_\ell^\pi} u_j \le B,
\]
where $q_i$ is the decode block allocated to request $i$ and $u_j$ is the prefill chunk allocated to request $j$.
Active decode requests are always served first, and prefill requests are admitted using the residual token budget.

\subsubsection{Chunk Continuation and KV-Safe Admission}
In our simulator, if a prefill request cannot be fully inserted into the current batch, it is chunked and its unfinished remainder is returned to the queue as a continuation request.
Continuation requests always have the highest admission priority among prefill candidates.
Let $M_{\ell}$ be the live KV-cache usage at batch boundary $\ell$.
Admission is checked against the safe threshold $M_{\rm safe}=\eta M_{\max}$, where $M_{\max}$ is the hard KV capacity, and $\eta \in (0,1)$ is a safety factor, reflecting the central role of KV-cache memory in LLM serving~\cite{kwon_pagedattention_2023}.
A candidate prefill request is admitted only if the projected KV usage remains below $M_{\rm safe}$.

\subsubsection{Batch Service Time}
Let $b_{\ell}$ be the total token load of batch $\ell$.
Its service time is modeled as $T_{\ell}=c_0+a\max(0,b_{\ell}-b_0)$, where $c_0$ is a fixed launch overhead, $a$ is the marginal per-token coefficient, and $b_0$ is a knee point obtained from profiling.

The event-driven simulator preserves these serving mechanics, including decode-priority batching, continuation handling, KV-safe admission, and token-load-dependent batch service times. Thus, all scheduling policies compared in this paper operate under the same token budget, KV-cache constraint, and batch-service model.

\subsection{Reduced \texorpdfstring{M/G/$n$}{M/G/n} Abstraction}

To obtain a compact analytical model that helps to interpret the key results, we collapse the two-stage execution of each request into a single effective service time, following standard queueing practice in which complex service mechanisms are represented by effective service-time distributions~\cite{harchol_balter_2013}:
\begin{equation}
S_i = s_0 + \alpha P_i + \beta D_i,
\label{eq:eff_service}
\end{equation}
where $s_0$ is a fixed overhead and $\alpha,\beta>0$ are effective prefill/decode coefficients.
Hereafter, $P$, $D$, and $S$ denote generic prompt-length, response-length, and effective-service-time random variables drawn from the same distribution as $P_i$, $D_i$, and $S_i$.
For discrete implementation,
\begin{equation}
S_i \asymp s_0
+ \tau_p \Big\lceil \frac{P_i}{C_p}\Big\rceil
+ \tau_d \Big\lceil \frac{D_i}{q}\Big\rceil,
\label{eq:eff_service_disc}
\end{equation}
where $C_p$ is the prefill chunk size, $q$ is the decode block size, and $\tau_p$ and $\tau_d$ are the effective service times of one prefill chunk and one decode block, respectively.

\subsubsection{Effective Parallelism}
Although the actual engine is not a classical multi-server queue, we approximate it as an M/G/$n$ system with effective parallelism $n=\lfloor \min(B/q,n_{\rm safe})\rfloor$, where $B/q$ is the decode concurrency implied by the token budget and $n_{\rm safe}$ is the maximum safe concurrency under KV constraints~\cite{harchol_balter_2013}.
Let $\mu=1/\mathbb E[S]$ and $\rho=\lambda \mathbb E[S]/n$.
The reduced system is stable when $\rho<1$.

\subsubsection{Moment Characterization}
From \eqref{eq:eff_service}, the first two moments are
\begin{align}
\mathbb E[S]
&= s_0+\alpha \mathbb E[P]+\beta \mathbb E[D], \\
\mathbb E[S^2]
&= s_0^2 + 2s_0\!\big(\alpha \mathbb E[P]+\beta \mathbb E[D]\big)
\nonumber\\
&\quad + \alpha^2 \mathbb E[P^2]
+ \beta^2 \mathbb E[D^2]
+ 2\alpha\beta \mathbb E[PD].
\end{align}
Let $C_s^2=\mathrm{Var}(S)/\mathbb E[S]^2$.
When arrivals are Poisson, $C_a^2=1$.

This abstraction collapses prefill,
decode, batching, and KV admission into a single effective service
time together with a static effective parallelism. It preserves
first-order mean-delay trends but cannot resolve token-level dynamics
such as decode-priority preemption, chunk continuation, or the
time-varying batch composition. We therefore use \eqref{eq:eff_service}--\eqref{eq:eff_service_disc}
only to explain mean-level trends, while tail latencies and
token-level metrics are evaluated in the simulator.

\subsection{Label Construction and Prediction Assumption}

To make the short/long partition well-defined, we consider that each request has an underlying response-length random variable $D_i$ drawn from a stationary workload distribution.
Rather than assuming that the predictor directly outputs an arbitrary binary label, we define the true short/long class through a workload-level cutoff.
Let $\tau_{0.85}$ be a fixed threshold chosen from the estimated response-length distribution, for example as a quantile corresponding to a desired short/long split.
The true class of request $i$ is defined as
\begin{equation}
Y_i = \begin{cases} S, & D_i\le\tau_{0.85}\\ L, & D_i>\tau_{0.85}\end{cases},
\label{eq:true_label_def}
\end{equation}

\subsubsection{Binary Predictor}
We assume a lightweight binary predictor that estimates the thresholded class $Y_i$ rather than the exact response length.
Accordingly, the scheduler observes only the predicted label $\hat Y_i \in \{S,L\}$, where $S$ and $L$ denote predicted-short and predicted-long requests.

\subsubsection{Prediction Errors}
The class priors are $\pi_S=\Pr(Y_i=S)$ and $\pi_L=1-\pi_S=\Pr(Y_i=L)$.
We characterize prediction errors by the two misclassification rates $\epsilon_{LS}=\Pr(\hat Y=S\mid Y=L)$ and $\epsilon_{SL}=\Pr(\hat Y=L\mid Y=S)$, which correspond to long-to-short and short-to-long errors, respectively.
Under this assumption, the predicted-class arrival rates satisfy
\begin{align}
\lambda_S
&=
\lambda\big[(1-\epsilon_{SL})\pi_S+\epsilon_{LS}\pi_L\big],\\
\lambda_L
&=
\lambda\big[\epsilon_{SL}\pi_S+(1-\epsilon_{LS})\pi_L\big].
\end{align}

\section{Scheduling Policies and Analytical Approximation}

This section describes the scheduling policies evaluated in the paper and their reduced analytical approximations. At each scheduling decision, active decode requests retain priority because they correspond to already-started generations. The policies below therefore mainly determine the admission order among queued prefill candidates after decode traffic and continuation requests have been handled.

\subsection{Baseline and Prediction-Guided Policies}

\subsubsection{FCFS Baseline}
First-come-first-served (FCFS) admits queued prefill requests in arrival order, subject to the token-budget and KV-cache constraints defined in the system model. Let $W_q$ denote queueing delay before service and $T$ denote end-to-end response time. For this baseline, we use the Allen--Cunneen approximation for a multi-server queue with general service-time variability~\cite{harchol_balter_2013}:
\begin{equation}
\mathbb E[W_q^{\rm FCFS}]
\approx
\frac{C_a^2+C_s^2}{2}
\cdot
\frac{P_{\rm wait}}{n\mu-\lambda},
\label{eq:fcfs_wq}
\end{equation}
where $P_{\rm wait}$ is the Erlang-C delay probability of the corresponding M/M/$n$ system.
Hence, $\mathbb E[T^{\rm FCFS}]\approx\mathbb E[S]+\mathbb E[W_q^{\rm FCFS}]$.

We model prediction-guided scheduling as a non-preemptive two-class priority discipline on top of the reduced M/G/$n$ abstraction.

\subsubsection{Predicted Classes}
Requests are partitioned into predicted-short and predicted-long classes.
Let $\lambda_S=\lambda \Pr(\hat Y=S)$ and $\lambda_L=\lambda \Pr(\hat Y=L)$, with $\lambda_S+\lambda_L=\lambda$.
Define the class-conditional moments $m_r^{(c)}=\mathbb E[S^r\mid \hat Y=c]$ for $c\in\{S,L\}$ and $r=1,2$, and the corresponding loads $\rho_c=\lambda_c m_1^{(c)}/n$.

\subsubsection{PSF Mean Waiting Approximation}
Predicted short-first (PSF) prioritizes predicted-short requests over predicted-long requests, while preserving arrival order within each predicted class, motivated by the classical mean-delay advantage of short-job-oriented service disciplines~\cite{schrage_srpt_1966}.
Exact closed forms for non-preemptive priority M/G/$n$ queues do not yield a compact expression.
We therefore use a compact priority correction that preserves the FCFS congestion term while capturing the short-first effect~\cite{harchol_balter_2013}.
Let $\Omega_n=(1-\rho)\mathbb E[W_q^{\rm FCFS}]$.
Then the queueing delays of the two predicted classes are approximated by
\begin{align}
\mathbb E[W_q^{(S)}]
&\approx
\frac{\Omega_n}{1-\rho_S},\\
\mathbb E[W_q^{(L)}]
&\approx
\frac{\Omega_n}{(1-\rho_S)(1-\rho)}.
\end{align}
The overall mean queueing delay under predicted short-first scheduling is
\begin{equation}
\mathbb E[W_q^{\rm PSF}]
\approx
\frac{\lambda_S}{\lambda}\mathbb E[W_q^{(S)}]
+
\frac{\lambda_L}{\lambda}\mathbb E[W_q^{(L)}],
\label{eq:psf_wq}
\end{equation}
and therefore $\mathbb E[T^{\rm PSF}]\approx\mathbb E[S]+\mathbb E[W_q^{\rm PSF}]$.

\subsubsection{Quantile-Based Aging}
To mitigate starvation of delayed predicted-long requests, the simulator uses a quantile-based aging threshold rather than a moment-based threshold, consistent with recent attention to queue-aware and scheduling-aware LLM serving~\cite{jiang_slob_2024}.
Let $W_{\rm ref}$ denote the empirical waiting-time sample used by the aging module.
The threshold is $\theta_A=Q_\gamma(W_{\rm ref})$, where $Q_\gamma(\cdot)$ is the empirical $\gamma$-quantile and $\gamma\in(0,1]$ is a tunable parameter.
A predicted-long request becomes \emph{aged} once its waiting time exceeds $\theta_A$.
The reference sample may be taken either from the current waiting queue or from the started-request history; in our experiments we use the history-based variant. 

\subsubsection{Aggressive Aging (AA)}
Aggressive Aging (AA) promotes all currently aged predicted-long requests ahead of ordinary predicted-short requests.
This mechanism serves as a natural fairness-oriented\footnote{In this paper, \emph{fairness} means tail-latency fairness, the protection of delayed long requests from starvation-induced tail inflation under short-first admission, rather than a Jain-index-style per-class balance or a direct short-versus-long latency gap. Accordingly, an \emph{efficiency--fairness trade-off} denotes the tension between reducing mean delay and controlling long-request tail latency.} extension of short-first scheduling: it mitigates starvation of delayed long requests, but can over-correct under heavy congestion because aged long requests may repeatedly occupy the short-first admission path.
This limitation motivates the bounded design of QBA.

\subsubsection{Quota-Based Aging (QBA)}
QBA combines a per-$k$-short-admissions quota with a promotion-debt
state to bound aged predicted-long promotions. A short-admission
counter advances \emph{only} when a short prefill fully completes
(not on chunk continuations); once it reaches $k$, one promotion
\emph{debt} is earned and cannot be displaced by further short
admissions, but debt is cleared (and the counter reset to 0)
\emph{only} when the oldest eligible aged-long request is
actually admitted; while debt is pending the counter stays at $k$,
serializing promotions to at most one per quota cycle.

\subsection{Mean-Level Surrogate and Simulator Alignment}

\subsubsection{Mean-Level Surrogate for AA and QBA}
In the analytical model, both AA and QBA are represented through effective promoted fractions estimated from simulator traces, yielding a reduced empirical closure rather than an exact priority-queue solution~\cite{harchol_balter_2013}.
For AA, let $\phi_A(\rho)$ denote the empirical fraction of predicted-long jobs whose waiting time exceeds the aging threshold at service start.
For QBA, let $\phi_Q(\rho;k)=N_{\rm promoted\text{-}long}/N_{\rm predicted\text{-}long}$, where the numerator counts successfully promoted predicted-long jobs under the quota rule and the denominator counts all predicted-long jobs.
These fractions are obtained at several load points and interpolated over $\rho$ to form load-dependent surrogates. 

The corresponding effective high-priority arrival rates are
\begin{align}
\lambda_S^{(A)} &= \lambda_S+\phi_A(\rho)\lambda_L,\\
\lambda_S^{(Q)} &= \lambda_S+\phi_Q(\rho;k)\lambda_L.
\end{align}
The promoted high-priority class then has mixed moments
\begin{equation}
\tilde m_r^{(S,X)}
=
\frac{\lambda_S m_r^{(S)}+\phi_X(\rho)\lambda_L m_r^{(L)}}
{\lambda_S+\phi_X(\rho)\lambda_L},
\qquad X\in\{A,Q\},\ r=1,2.
\label{eq:mixed_moments_promoted}
\end{equation}
Here $\phi_X(\rho)$ denotes $\phi_A(\rho)$ for $X=A$ and $\phi_Q(\rho;k)$ for $X=Q$.
Substituting these mixed moments into the PSF approximation yields reduced mean-level surrogates for the aggressive-aging variant AA and the bounded policy QBA.
For this closure, $\lambda_L^{(X)}=\lambda-\lambda_S^{(X)}$, $\rho_S^{(X)}=\lambda_S^{(X)}\tilde m_1^{(S,X)}/n$, and $\mathbb E[W_q^{(S,X)}]$ and $\mathbb E[W_q^{(L,X)}]$ are obtained by applying the same priority-correction form as in \eqref{eq:psf_wq} with the mixed promoted class.
Thus, $\phi_X(\rho)$ controls the effective high-priority population: larger promoted fractions improve delayed-long access but also increase congestion in the short-first path.
\begin{equation}
\mathbb E[W_q^{X}]
\approx
\frac{\lambda_S^{(X)}}{\lambda}\mathbb E[W_q^{(S,X)}]
+
\frac{\lambda_L^{(X)}}{\lambda}\mathbb E[W_q^{(L,X)}],
\qquad X\in\{A,Q\}.
\label{eq:ewq_AA_QBA}
\end{equation}

The reduced closures in~\eqref{eq:ewq_AA_QBA} and the simulator share
the same workload family and prediction setup; the analytical layer
collapses engine details for mean-level interpretation, while the
simulator preserves token-level dynamics for tail-latency, TTFT, and
fairness-efficiency evaluation.

\begin{figure*}[t]
\centering
\subfloat[{Formula Mean waiting time $E[W_q]$}]{%
  \includegraphics[width=0.32\textwidth]{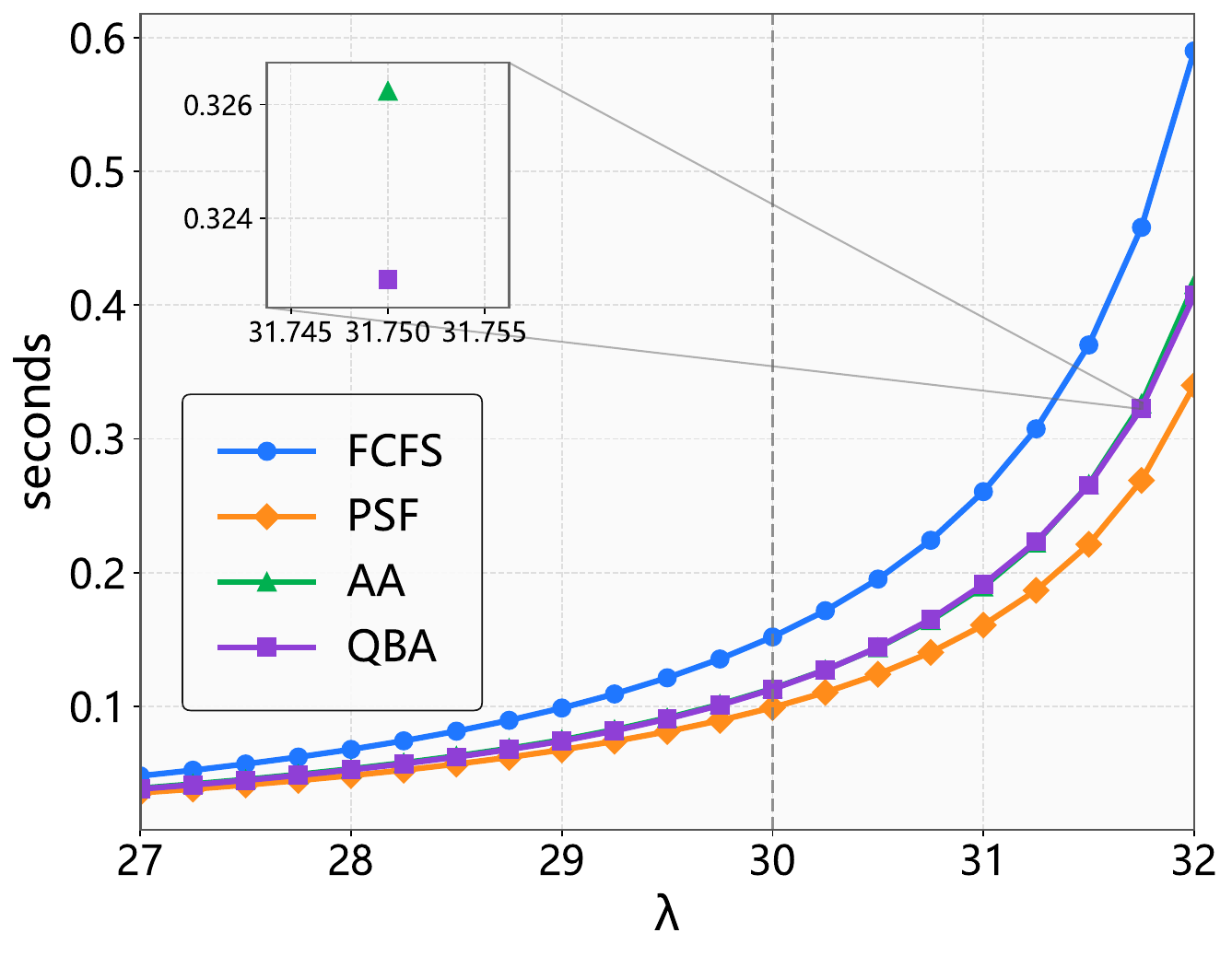}%
  \label{subfig:Ana_mean_wait}%
}%
\subfloat[Simulation Mean waiting time]{%
  \includegraphics[width=0.32\textwidth]{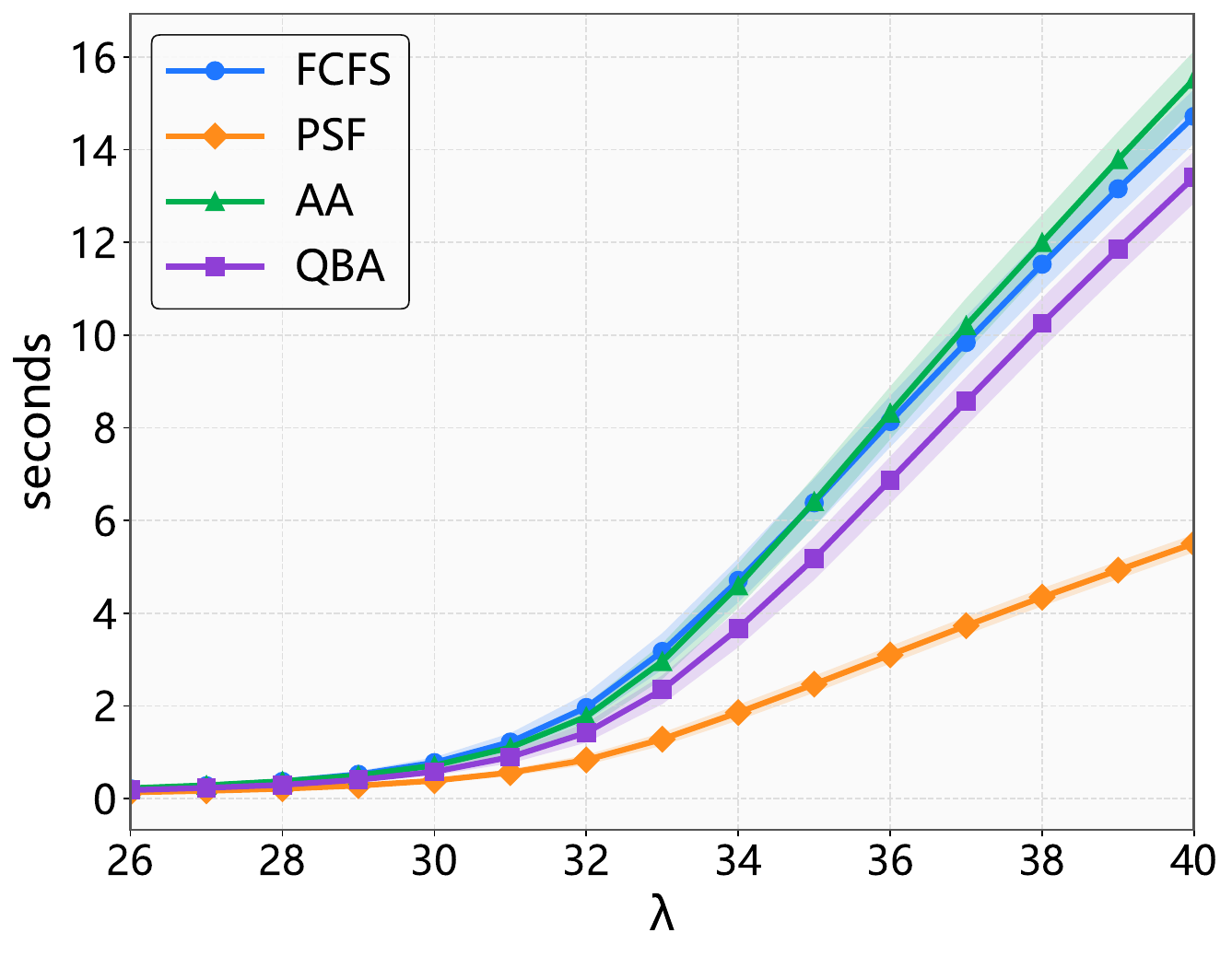}%
  \label{subfig:Sim_mean_wait}%
}%
\subfloat[Simulation P99 waiting time]{%
  \includegraphics[width=0.32\textwidth]{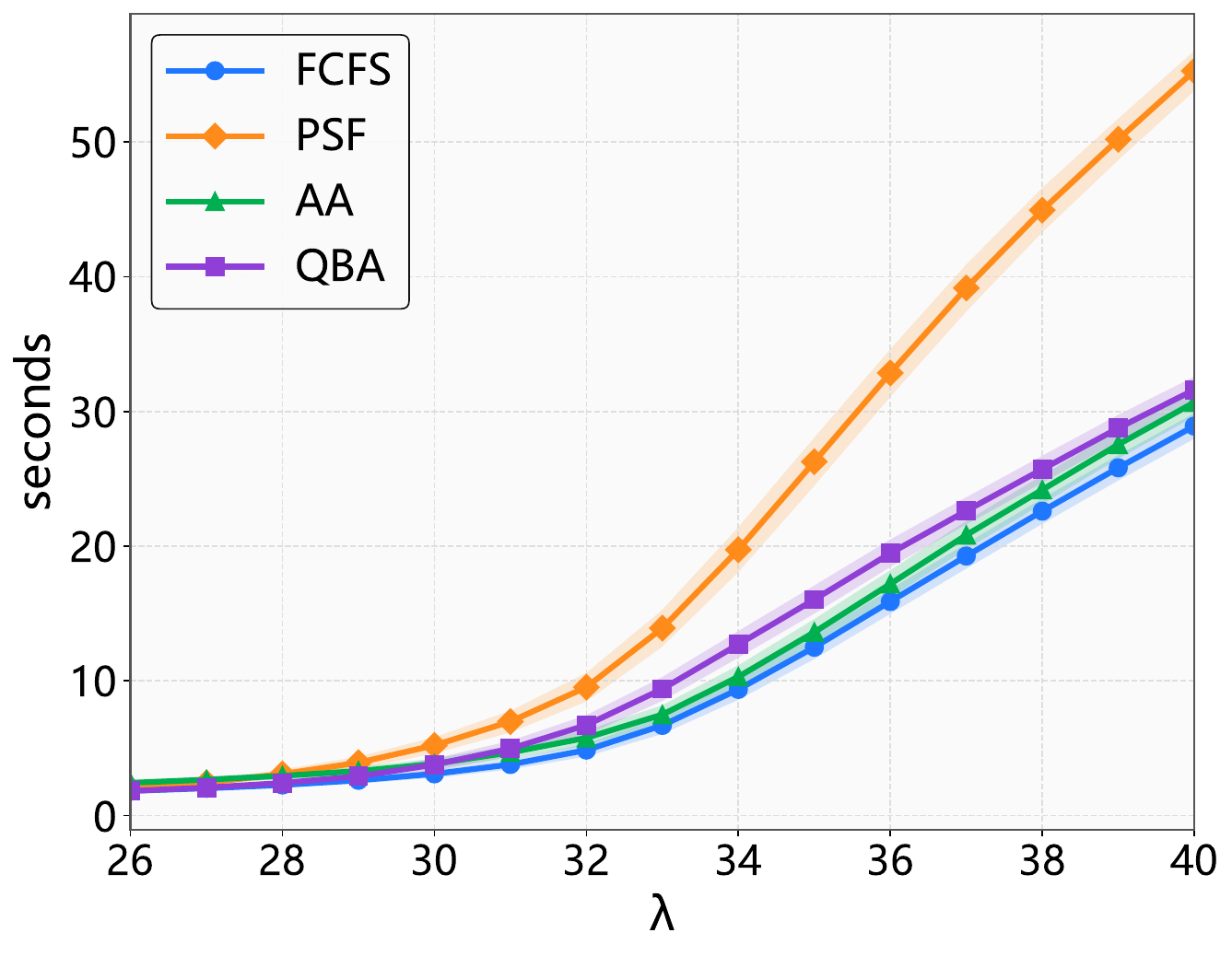}%
  \label{subfig:Sim_p99_wait}%
}
\caption{Mean and tail waiting-time comparison of FCFS, PSF, AA, and QBA under the reduced analytical model and the high-fidelity simulator.}
\label{fig:base_comparison}
\end{figure*}

\section{Experiments}

\subsection{Experimental Setup}

\subsubsection{Synthetic Workload Generation}
We evaluate all policies on a synthetic request stream designed to
mimic heterogeneous LLM serving workloads~\cite{xiang_servegen_2025}.
Requests arrive according to a Poisson process with arrival rate
$\lambda$. For each request, the prompt length is sampled from a
mixture distribution: with probability $0.7$ it is drawn from
$\mathcal{N}(100,30^2)$, and with probability $0.3$ it is drawn from
a Pareto distribution with shape $1.4$ and scale $40$, reflecting the
heavy-tailed prompt-length behavior observed in production LLM
workloads~\cite{xiang_servegen_2025}. Both prompt and response lengths
are capped at $4096$ tokens. Conditioned on the prompt length $P_i$,
the response length $D_i$ is generated from a lognormal model
\[
D_i \sim
\mathrm{LogNormal}\!\big(\beta_0+\beta_1\log(1+P_i),\,\sigma^2\big),
\]
with $\beta_0=3$, $\beta_1=0.75$, and $\sigma=0.3$.

\subsubsection{Prediction Labels}
We construct short/long labels using the empirical $0.85$-quantile
of the generated response-length distribution. Unless otherwise stated,
the predicted label matches the thresholded ground-truth label exactly.
This noiseless setting isolates the scheduling mechanism; asymmetric
prediction errors are evaluated separately below.

\subsubsection{Serving-System Configuration}
The serving engine uses a token budget of $B=128$ per iteration and
a decode block size of $q=8$ tokens. The KV-cache capacity is set to
$8192$ tokens, and admission is guarded by a safety threshold equal
to $0.7$ of the total KV budget. Batch service time follows the
profiled piecewise-linear model with parameters $c_0=0.003$,
$a=2\times10^{-5}$, and knee point $b_0=64$; concurrency correction
and timing jitter are disabled in the main experiments. For the
aging-based schedulers, the aging module uses a history-based reference
sample with quantile parameter $\gamma=0.80$. QBA additionally uses
quota $k=5$. These values are used as the default configuration unless
the corresponding parameter is explicitly varied in a sensitivity
experiment.

\subsubsection{Policy Comparison Protocol}
We compare four schedulers defined in Section~III:
\textbf{FCFS}, \textbf{PSF}, \textbf{AA}, and \textbf{QBA}.
All policies operate under the same token budget, KV-cache constraint,
prediction labels, and batch-service model. For the main load-sweep
experiment, we vary the arrival rate over
$\lambda \in \{26,27,28,\dots,40\}$ and report the mean waiting time
together with the $99$th percentile waiting time. Each simulated
operating point is repeated over $100$ random seeds. We report
seed-averaged results, and shaded bands indicate the corresponding
confidence intervals.

\subsubsection{Analytical Protocol}
The reduced analytical model is calibrated from an FCFS trace at the
calibration load $\lambda=30$. This calibration yields the effective
service moments and the effective parallelism parameter used in the
M/G/$n$ approximation. The calibration point is chosen in the congested regime so that the
effective service moments reflect a non-trivial queue, and it differs
from the $\lambda=32$ reference load used for the main comparisons. For AA and QBA, the effective promoted fractions
are not assumed a priori; instead, they are estimated from simulator
traces at $\lambda \in \{25,27,29,31,33\}$ and interpolated as
load-dependent surrogate functions in the reduced model. The objective
of the analytical comparison is not exact numerical matching with the
simulator, but a reduced mean-level characterization of the scheduling
trends.

\subsubsection{Prediction-Error Protocol}
To evaluate sensitivity to imperfect length prediction, we consider
two asymmetric label errors. We define the short-to-long error rate as
\[
\epsilon_{SL}
=
\Pr(\hat{Y}=\mathrm{long}\mid Y=\mathrm{short}),
\]
and the long-to-short error rate as
\[
\epsilon_{LS}
=
\Pr(\hat{Y}=\mathrm{short}\mid Y=\mathrm{long}),
\]
where $Y$ and $\hat{Y}$ denote the ground-truth and predicted request
classes, respectively. The two error directions are evaluated
independently: we sweep one error rate over
$\{0,0.05,0.10,\ldots,0.40\}$ while fixing the other to zero.
For each error level, the same scheduling policies and serving-system
configuration are used as in the main comparison.

Each operating point is repeated over $100$ paired random seeds.
In addition to overall mean waiting time, we report the $99$th
percentile waiting time separately for the true-short and true-long
request classes. These class-conditional metrics are used to reveal
latency redistribution that may be obscured by aggregate averages,
particularly when prediction errors affect the two request classes
asymmetrically.

\begin{figure*}[t]
\centering
\subfloat[Simulation Mean waiting time vs quota k]{%
  \includegraphics[width=0.32\textwidth]{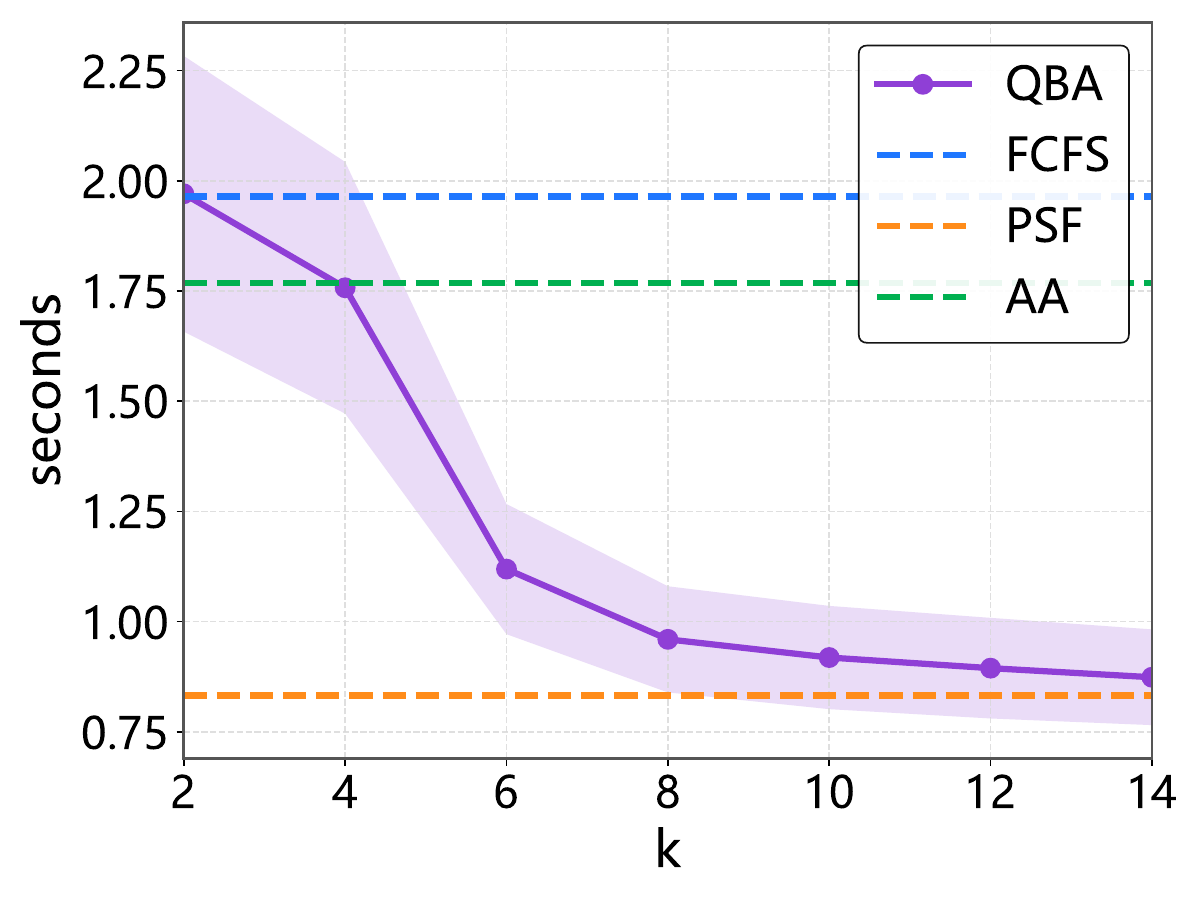}%
  \label{subfig:quota_mean}%
}%
\subfloat[Simulation P99 waiting time vs quota k]{%
  \includegraphics[width=0.32\textwidth]{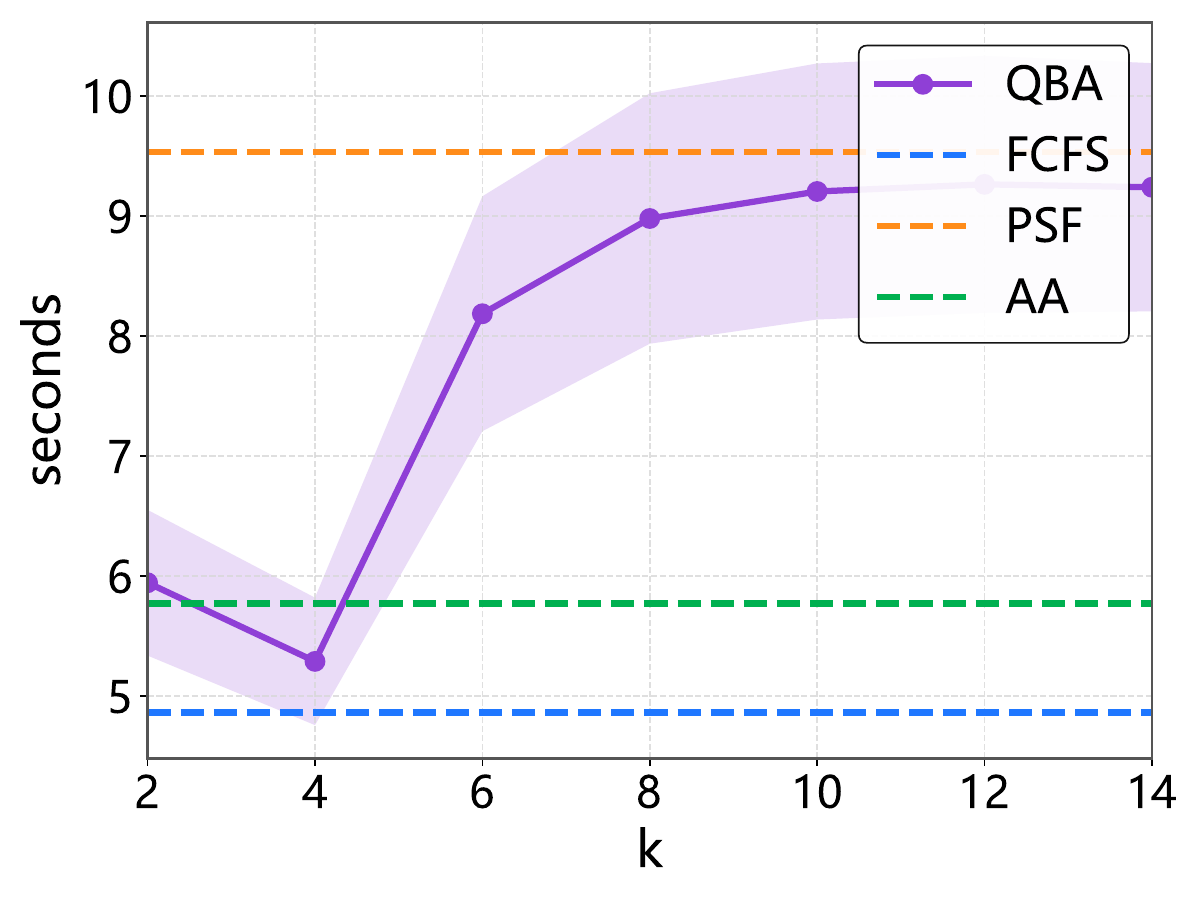}%
  \label{subfig:quota_p99}%
}%
\subfloat[{Simulation P99 waiting time vs $\gamma$}]{%
  \includegraphics[width=0.32\textwidth]{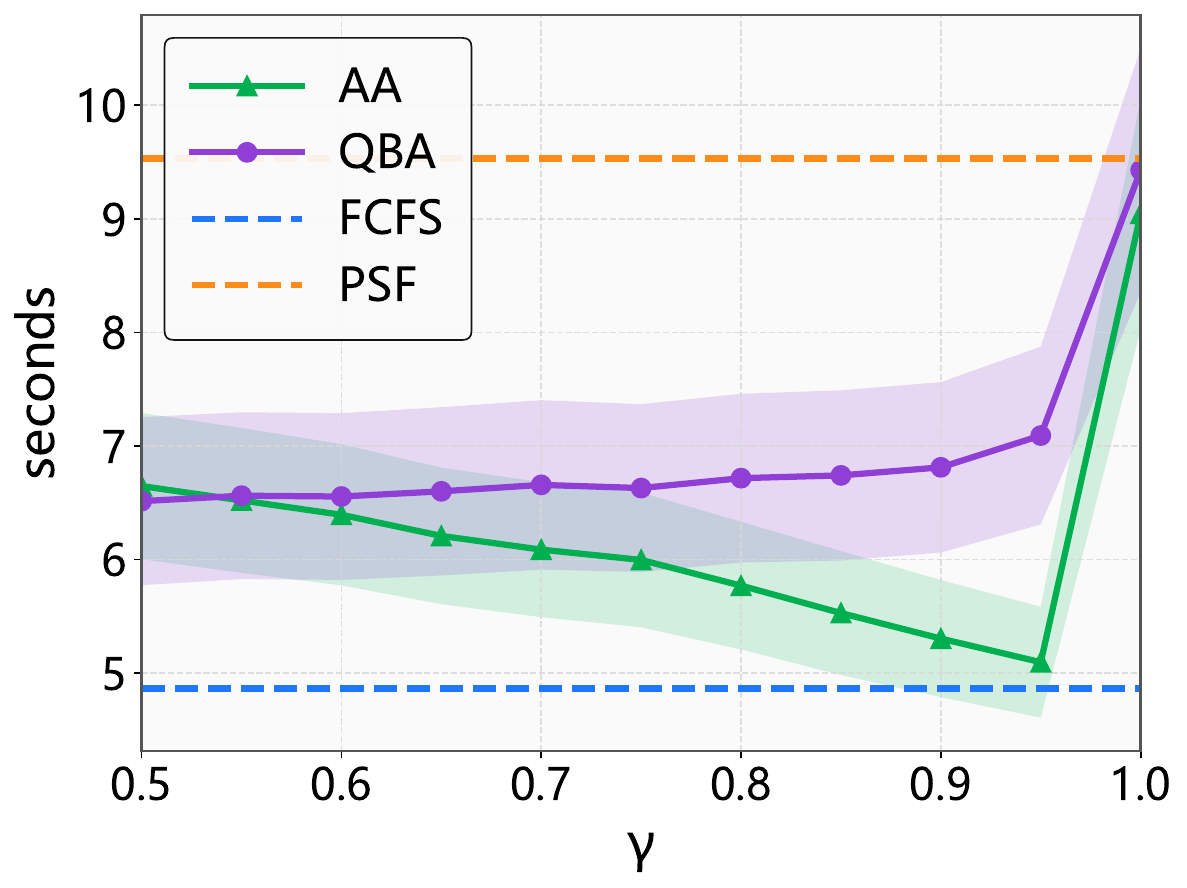}%
  \label{subfig:gamma_p99}%
}
\caption{Sensitivity of QBA to the quota parameter $k$ and the aging-quantile-level parameter $\gamma$ at the reference load. Solid curves show seed-averaged simulation results over 100 random seeds, and shaded bands indicate confidence intervals. Dashed horizontal lines provide baseline references for context.}
\label{fig:QBA_sensitivity}
\end{figure*}

\subsection{Performance Comparison}

Fig.~\ref{fig:base_comparison} compares FCFS, PSF, AA, and QBA using both the reduced analytical model and the high-fidelity simulator.

We first examine the mean-level trends predicted by the analytical model in Fig.~\ref{fig:base_comparison}\subref{subfig:Ana_mean_wait}.
The reduced model yields a consistent first-order ordering: all prediction-guided policies improve over FCFS in mean waiting time, PSF achieves the lowest mean delay, and both AA and QBA remain between PSF and FCFS over the evaluated load range.
Moreover, QBA remains very close to AA in the analytical curves, and is slightly lower over most of the evaluated range, indicating that quota-based bounded aging preserves most of the benefit of aging while avoiding overly aggressive promotion of delayed long requests.

The simulator in Fig.~\ref{fig:base_comparison}\subref{subfig:Sim_mean_wait} reveals a more realistic trade-off.
PSF consistently minimizes the mean waiting time, confirming the efficiency advantage of short-first scheduling.
However, the aggressive aging policy AA over-corrects under congestion and eventually exceeds FCFS in mean delay.
In contrast, QBA substantially mitigates this degradation: across the tested loads, its mean waiting time remains below FCFS and significantly below AA, while preserving a substantial fraction of the efficiency advantage of PSF.
This result supports the main motivation of QBA: bounded and enforceable promotion preserves mean-delay performance more reliably than unconstrained aging.

The tail behavior is shown in Fig.~\ref{fig:base_comparison}\subref{subfig:Sim_p99_wait}.
As expected, PSF exhibits the most severe tail inflation, especially at high load, confirming that pure short-first prioritization amplifies starvation risk for delayed long requests.
Both aging-based policies substantially reduce this extreme-tail growth relative to PSF.
However, the seed-averaged results show a more nuanced ordering than in the single-run case.
At low load, FCFS and QBA are very close in the tail, and both remain well below PSF.
Under moderate and high load, FCFS becomes the strongest tail baseline, AA attains a lower P99 than QBA, and QBA still remains far below PSF.
Accordingly, the main advantage of QBA is a more favorable mean--tail trade-off than aggressive aging when average delay is also an important objective.

Overall, Fig.~\ref{fig:base_comparison} still validates the basic scheduling picture studied in this paper: PSF is the most efficient but least fair baseline, AA represents an aggressive fairness-oriented correction that sacrifices mean delay, and QBA serves as a practical middle ground that recovers much of the mean-delay advantage of prediction-guided scheduling while avoiding the extreme tail inflation of pure short-first prioritization.

\begin{figure*}[t]
\centering
\subfloat[Overall mean waiting time]{%
  \includegraphics[width=0.32\textwidth]{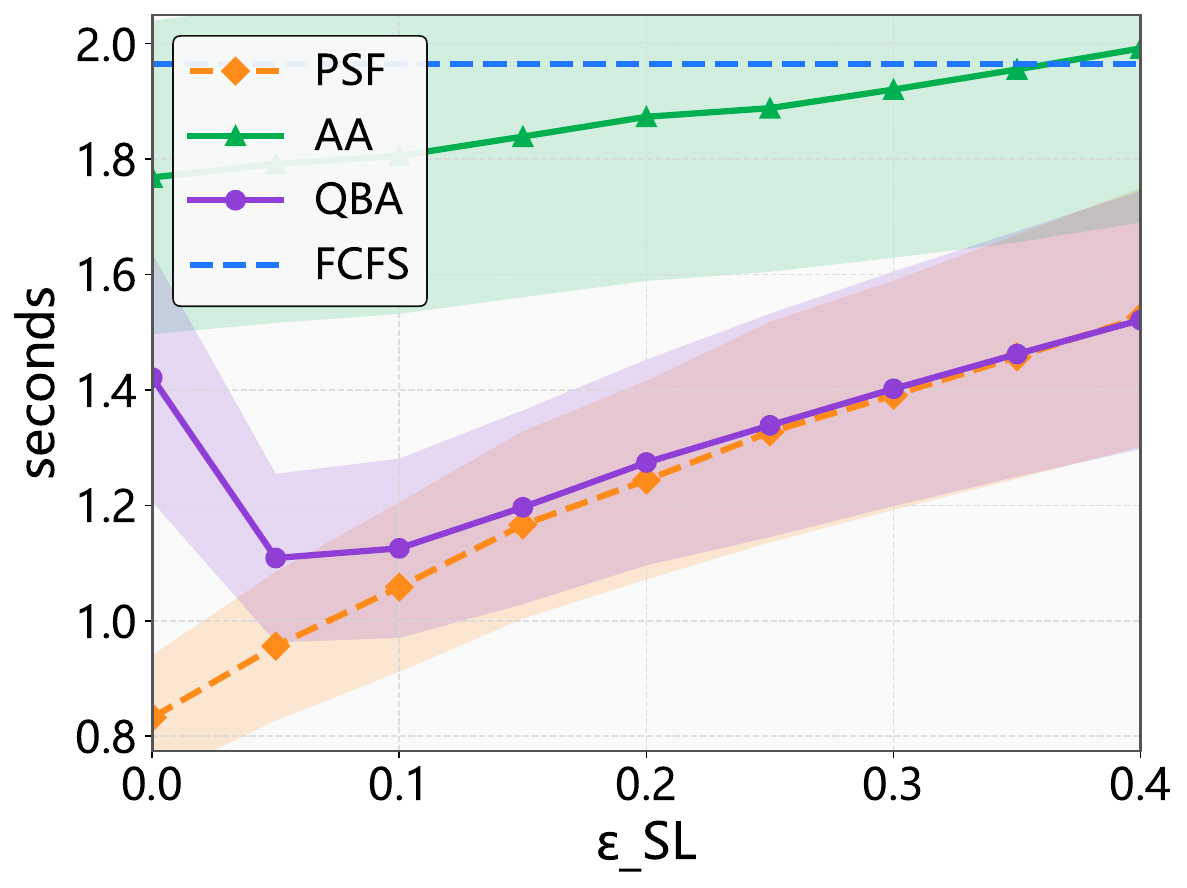}%
  \label{subfig:epsilonSL_mean_waiting}%
}%
\subfloat[True-short P99 waiting time]{%
  \includegraphics[width=0.32\textwidth]{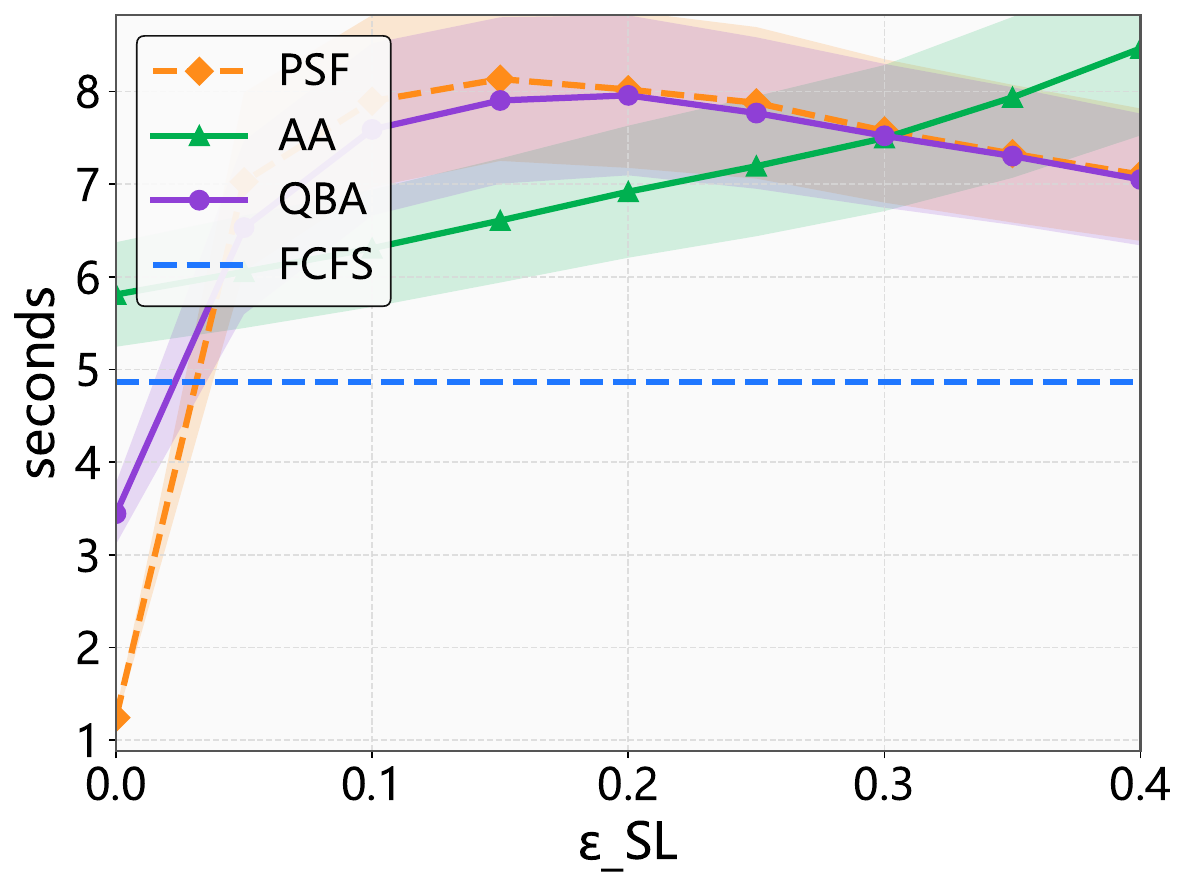}%
  \label{subfig:epsilonSL_true_short_p99}%
}%
\subfloat[True-long P99 waiting time]{%
  \includegraphics[width=0.32\textwidth]{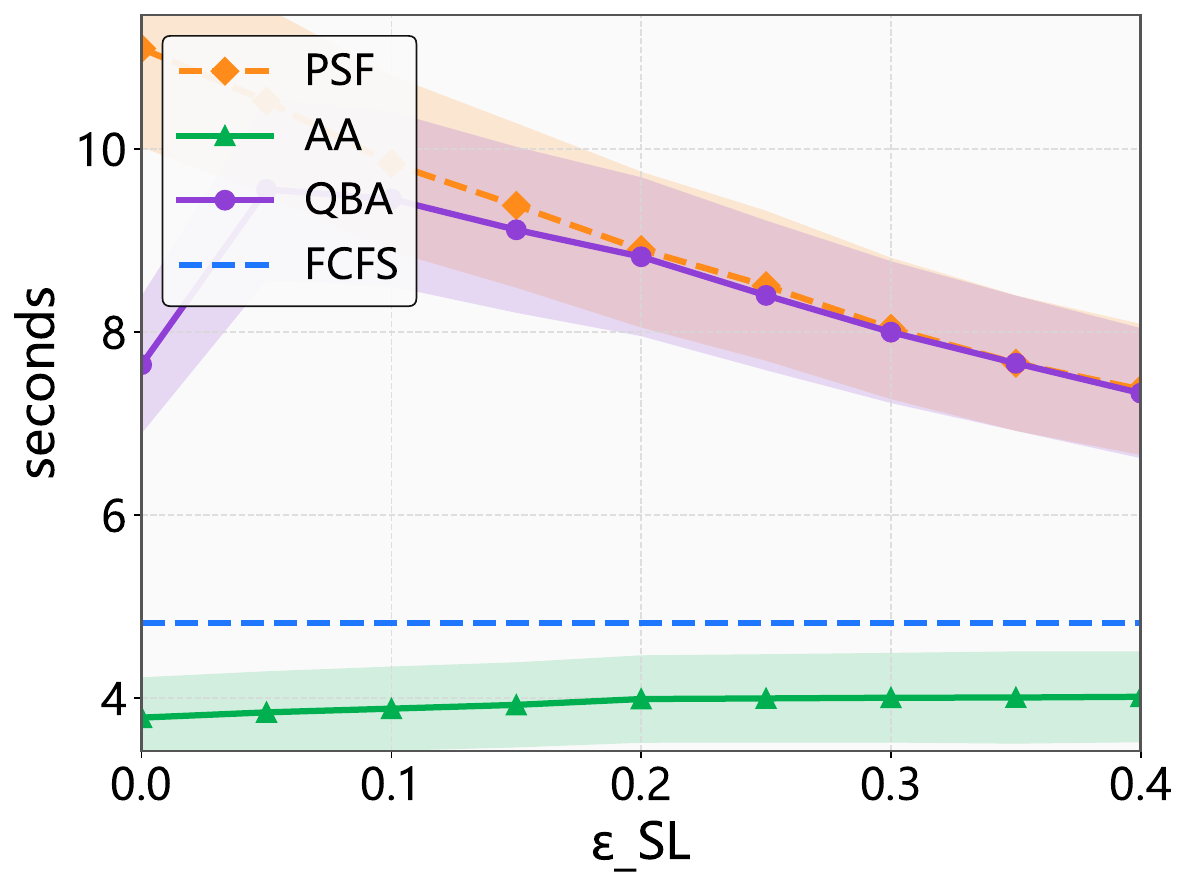}%
  \label{subfig:epsilonSL_true_long_p99}%
}
\caption{Sensitivity to short-to-long prediction errors ($\epsilon_{LS}=0$, $\lambda=32$).
Panels: mean, true-short P99, true-long P99.
Same 100-seed convention as Fig.~\ref{fig:QBA_sensitivity}.}
\label{fig:epsilonSL_analysis}
\end{figure*}

\subsection{Sensitivity Analysis of QBA Hyperparameters}

We next study the sensitivity of QBA to its two main hyperparameters: the quota parameter $k$ and the aging-quantile-level parameter $\gamma$.
All results in Fig.~\ref{fig:QBA_sensitivity} are obtained at the reference load $\lambda=32$.
For the quota sweep, we vary $k$ from $2$ to $14$ with step size $2$ while fixing $\gamma=0.8$.
For the aging sweep, we vary $\gamma$ from $0.5$ to $1.0$ with step size $0.05$ while fixing $k=5$.

The first two panels of Fig.~\ref{fig:QBA_sensitivity} show that the quota parameter $k$ acts as the primary control knob of QBA.
As shown in Fig.~\ref{fig:QBA_sensitivity}\subref{subfig:quota_mean}, the mean waiting time is nearly flat for very small $k$ and then decreases markedly once $k$ reaches a moderate range.
This indicates that reducing the frequency of long-request promotion gradually restores more of the efficiency advantage of short-first scheduling.
In contrast, Fig.~\ref{fig:QBA_sensitivity}\subref{subfig:quota_p99} reveals a non-monotonic tail trend: the P99 waiting time is minimized at a moderate value around $k=4$, and then increases as $k$ becomes larger.
This implies that the quota parameter does not merely interpolate linearly between fairness and efficiency.
Instead, it creates a useful operating region: small $k$ behaves more like aggressive aging, large $k$ drifts toward PSF, and an intermediate $k$ yields the best tail control within the QBA family while retaining a substantial mean-delay gain. In particular, the lowest P99 within the QBA family is attained around $k=4$, although FCFS still remains the lowest-tail baseline at the same reference load.

The third panel, Fig.~\ref{fig:QBA_sensitivity}\subref{subfig:gamma_p99}, shows that $\gamma$ functions mainly as a secondary trigger-strength parameter.
For QBA, the P99 waiting time remains relatively flat over a broad interval from approximately $\gamma=0.5$ to $\gamma=0.95$, and deteriorates sharply only when $\gamma$ approaches $1.0$.
By contrast, AA is more sensitive to $\gamma$: its P99 improves steadily as $\gamma$ increases toward $0.95$, but also collapses at $\gamma=1.0$ when aging becomes effectively disabled.
Therefore, the role of $\gamma$ is different from that of $k$.
The quota parameter primarily determines the policy's position on the efficiency-fairness spectrum, whereas $\gamma$ mostly fine-tunes the aggressiveness of aging around that operating point.
In this sense, QBA is not the lowest-tail policy under $\gamma$ tuning, but it is the more robust one across a wide range of threshold settings.

\begin{figure*}[t]
\centering
\subfloat[Overall mean waiting time]{%
  \includegraphics[width=0.32\textwidth]{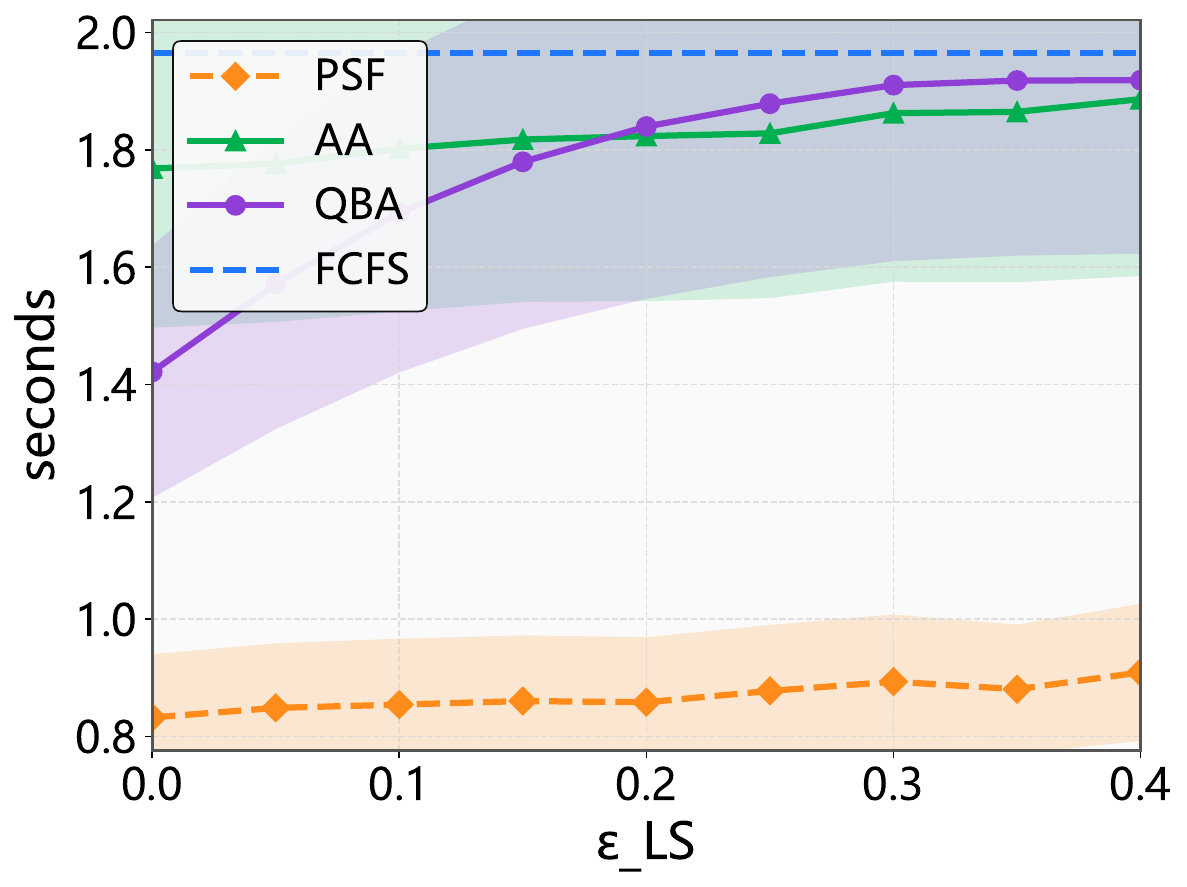}%
  \label{subfig:epsilonLS_mean_waiting}%
}%
\subfloat[True-short P99 waiting time]{%
  \includegraphics[width=0.32\textwidth]{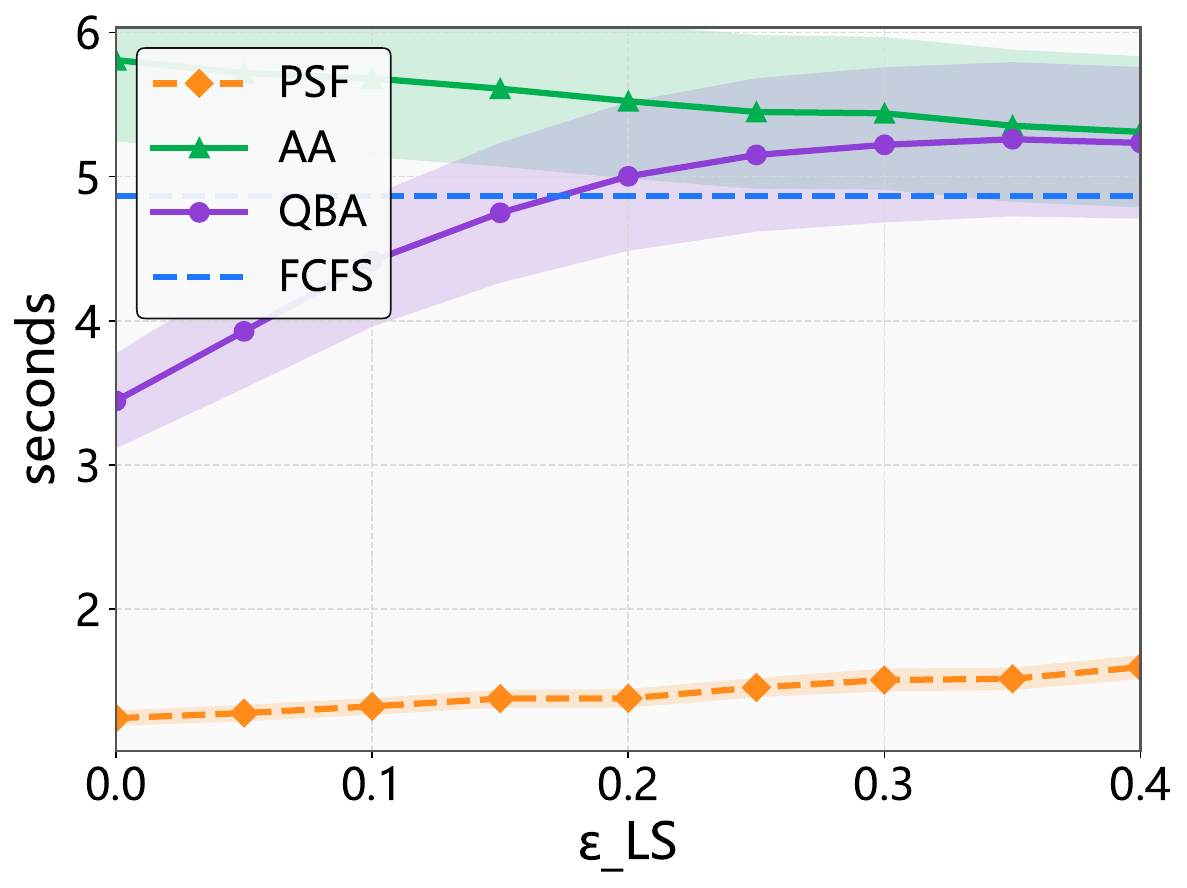}%
  \label{subfig:epsilonLS_true_short_p99}%
}%
\subfloat[True-long P99 waiting time]{%
  \includegraphics[width=0.32\textwidth]{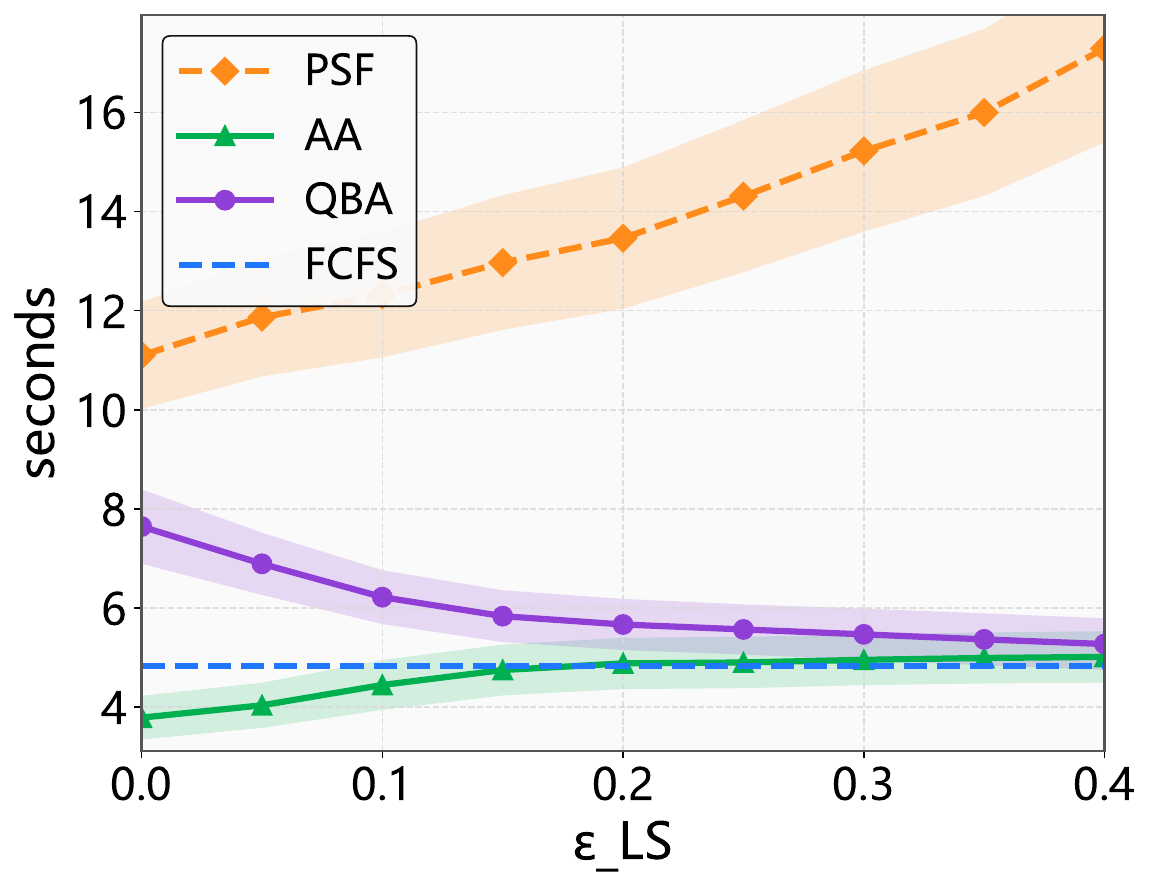}%
  \label{subfig:epsilonLS_true_long_p99}%
}
\caption{Sensitivity to long-to-short prediction errors ($\epsilon_{SL}=0$, $\lambda=32$).
Panels: mean, true-short P99, true-long P99.
Same 100-seed convention as Fig.~\ref{fig:QBA_sensitivity}.}
\label{fig:epsilonLS_analysis}
\end{figure*}

\subsection{Sensitivity to Asymmetric Prediction Errors}

Figure~\ref{fig:epsilonSL_analysis} first demonstrates the results of short-to-long
errors, where genuinely short requests are incorrectly assigned to
the predicted-long class. For PSF, increasing $\epsilon_{SL}$
progressively removes short requests from the high-priority class,
raising the overall mean waiting time. More importantly, even a small
error rate sharply increases the true-short tail: at
$\epsilon_{SL}=0.05$, the P99 waiting time of true-short requests
increases from approximately $1.24$~s to $7.03$~s. This demonstrates
that aggregate mean latency alone does not capture the impact of
prediction errors on the requests that lose their intended priority.

QBA exhibits a different response to short-to-long errors. Its mean
waiting time initially decreases, reaching approximately $1.13$~s at
$\epsilon_{SL}=0.10$, but this apparent improvement is accompanied by
substantial degradation in both class-conditional tails. Over the same
range, the true-short P99 increases from $3.44$~s to $7.59$~s and the
true-long P99 increases from $7.64$~s to $9.45$~s. Hence, the mean-delay reduction is not increased robustness, but a redistribution across true classes.

This behavior is consistent with contamination of the predicted-long
class: short requests that are misclassified as long enter the same
population from which aging-based promotions are selected, potentially
diluting the protection intended for genuinely long requests. As
$\epsilon_{SL}$ becomes large, PSF and QBA gradually converge because
the predicted classes themselves carry progressively less useful
length information.

Figure~\ref{fig:epsilonLS_analysis} shows a qualitatively different
failure mode under long-to-short errors. PSF remains highly efficient
in terms of the aggregate mean: increasing $\epsilon_{LS}$ from $0$
to $0.4$ changes its mean waiting time only from approximately
$0.83$~s to $0.91$~s. Its true-short P99 also increases only
moderately. These aggregate and short-class metrics, however, conceal
a substantial deterioration in the long-request tail.

The true-long P99 of PSF rises from approximately $11.10$~s under
perfect prediction to $17.29$~s at $\epsilon_{LS}=0.4$, an increase
of about $56\%$. Under QBA, increasing $\epsilon_{LS}$ redistributes
latency between the two true classes: the true-long P99 falls from
$7.64$~s to $5.27$~s, while the true-short P99 rises from $3.44$~s
to $5.23$~s as misclassified long requests enter the predicted-short
path. At $\epsilon_{LS}=0.4$, QBA still cuts the true-long P99 by
about $70\%$ versus PSF, bounding the long-tail amplification that pure PSF suffers.

This tail protection is not free. QBA's overall mean waiting time
increases from approximately $1.42$~s to $1.92$~s as
$\epsilon_{LS}$ increases to $0.4$, leaving only a marginal
mean-latency advantage over FCFS at high error rates. Its true-short
P99 also increases from $3.44$~s to $5.23$~s. Therefore, QBA should
not be interpreted as prediction-error invariant: long-to-short errors
primarily erode its efficiency advantage while shifting latency toward
the true-long class, rather than leaving its tail protection stable.
AA shows a similarly stable tail response at high $\epsilon_{LS}$,
indicating that aging-based fairness mechanisms in general are less
susceptible than pure PSF to long-tail amplification in this error
direction. QBA's distinguishing role is therefore its quota-controlled
intervention under accurate or moderately accurate prediction, rather than uniform dominance over unrestricted aging.

Taken together, Figs.~\ref{fig:epsilonSL_analysis} and
\ref{fig:epsilonLS_analysis} show that prediction errors cannot be
characterized by a single accuracy value. Their impact depends both
on the direction of misclassification and on the scheduling mechanism.
Short-to-long errors degrade class separation and can interfere with
QBA's targeting of delayed predicted-long requests, whereas
long-to-short errors progressively transfer priority toward true-long
requests and erode the short-first efficiency advantage. For pure PSF,
the latter error direction is particularly harmful to true-long tail
latency despite only a small change in the aggregate mean. QBA
therefore does not eliminate prediction sensitivity; instead, it
changes the failure mode of prediction-guided short-first scheduling
and, in both error directions, avoids the severe long-tail amplification
exhibited by pure PSF.

\subsection{Production-Trace Validation}
We next test whether the same efficiency--fairness trade-off
extends from synthetic workloads to a 100-window Microsoft Azure
LLM trace~\cite{patel_splitwise_2024} with no change to the engine
or scheduler. Each window preserves original request ordering and relative
inter-arrival structure, with the timeline normalized to
$\lambda=32$; prompt and response lengths are uniformly scaled by
a factor of $0.44$ to match synthetic demand. All serving-system and scheduling
parameters are kept unchanged. The resulting traces remain strongly
non-Poisson, with a mean inter-arrival squared coefficient of variation
of approximately $2.9$.

Table~\ref{tab:azure_trace} shows that the production traces preserve
the efficiency--fairness tension observed in the controlled
experiments. PSF reduces mean waiting time by $31\%$ relative to FCFS,
but increases true-long P99 from $0.77$~s to $2.10$~s. QBA reduces
this long-tail value to $1.45$~s while retaining a $18\%$ mean-delay
improvement over FCFS. Across all 100 trace windows, PSF, QBA, and AA
consistently form the same ordering in mean delay, whereas the ordering
is reversed for true-long P99. Thus, without workload-specific
retuning, QBA remains a stable intermediate operating point between
short-first efficiency and aggressive aging-based tail protection. Throughput is essentially identical across all four policies
($\approx 2.47\times10^4$ tokens/s), batch utilization within
0.89--0.91, with zero OOM events across all windows.

\begin{table}[t]
\centering
\caption{Production-trace-driven evaluation over 100 non-overlapping
Microsoft Azure LLM inference windows.
Reported P99s are per-window medians; mean wait is pooled
across all requests.}
\label{tab:azure_trace}
\begin{tabular}{lccccc}
\hline
Policy & Mean Wait & P99 Wait & Short P99 & Long P99 & TTFT P99 \\
       & (s) & (s) & (s) & (s) & (s) \\
\hline
FCFS & 0.179 & 0.771 & 0.771 & 0.766 & 0.804 \\
PSF  & \textbf{0.123} & 1.376 & \textbf{0.285} & 2.097 & 1.428 \\
AA   & 0.178 & 0.949 & 0.970 & \textbf{0.491} & 0.975 \\
QBA  & 0.146 & 0.951 & 0.565 & 1.446 & 0.998 \\
\hline
\end{tabular}
\end{table}

\section{Conclusion}
QBA is a bounded-aging scheduler for token-level continuous batching
in LLM serving, regulating aged-long promotions through a
per-$k$-short-admissions quota with enforceable promotion debt.
Validated on FCFS/PSF/AA over synthetic and 100-window Azure traces,
QBA retains roughly $59\%$ of PSF's mean-delay gain while
substantially reducing tail inflation, with no aged predicted-long
request indefinitely postponable under bounded $k$.

\section*{Acknowledgment}
This work was partly supported by the Guangdong Provincial Key
Laboratory IRADS, BNBU (Project code 2022B1212010006), and the
Guangdong Higher Education Upgrading Plan (2021--2025) UIC
R0400001-22 and R0400024-22.


%





\ifCLASSOPTIONcaptionsoff
  \newpage
\fi

\balance
\bibliographystyle{IEEEtran}
\bibliography{IEEEabrv,Bibliography}


\end{document}